\documentclass{moriond}

\newcommand{\dd}[0]{\ensuremath{\mathrm{d}}}
\usepackage{amsmath,amssymb}
\usepackage{graphicx}
\DeclareGraphicsExtensions{.pdf, .png, .jpg}
\graphicspath{{figures/}}

\newcommand{\Photo}{\includegraphics[height=35mm]{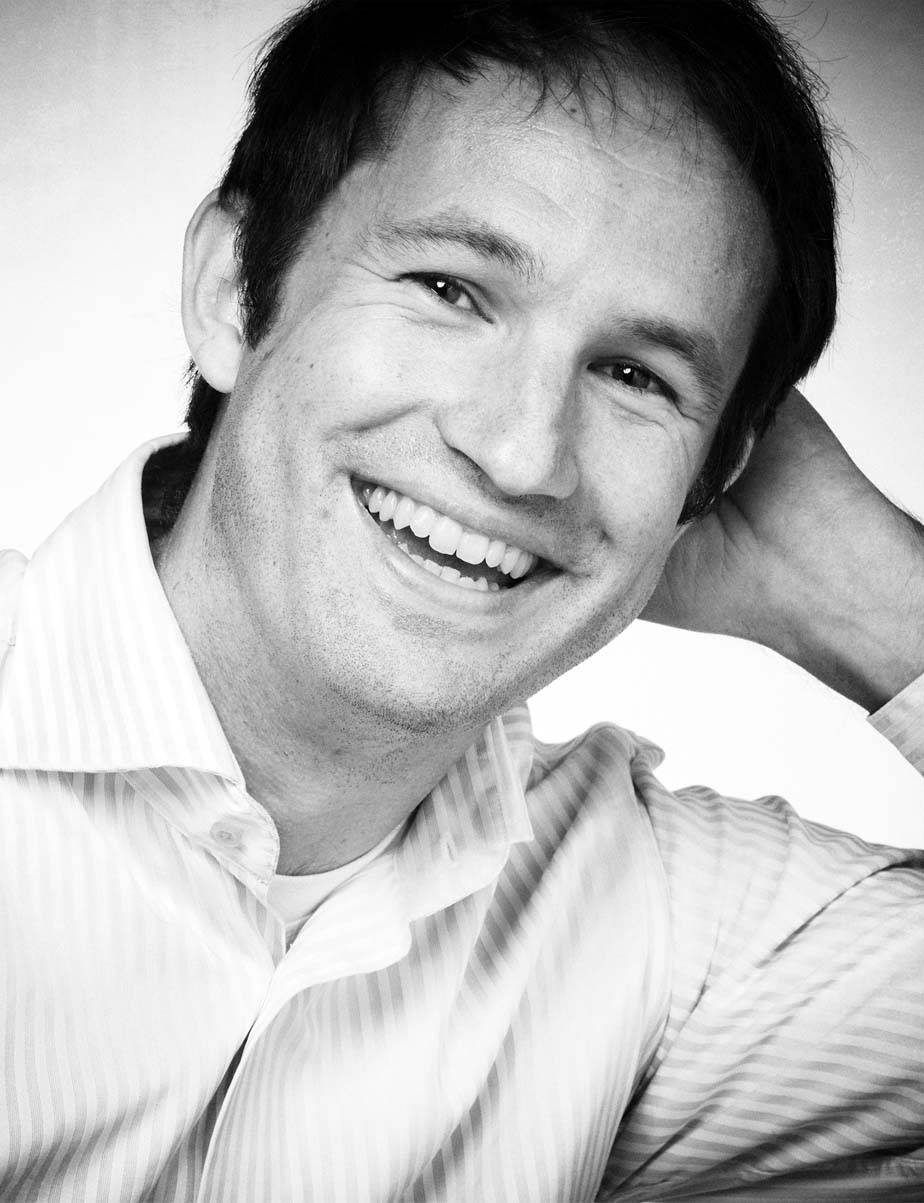}}

\begin{document}
\vspace*{4cm}
\title{Cosmic ray electron and positron spectra at TeV energies}

\author{P. MERTSCH}

\address{Institute for Theoretical Particle Physics and Cosmology (TTK), RWTH Aachen University, Sommerfeldstr. 16, 52074 Aachen, Germany}

\maketitle\abstracts{
Observations of cosmic ray electrons have made great strides in the last decade and direct observations of the all-electron flux as well as separate electron and positron spectra are now available up to $\sim 1 \, \text{TeV}$. In this invited contribution to the 2022 edition of the \textit{Rencontres de Moriond} on ``Very High Energy Phenomena in the Universe'', we review the data on cosmic ray electron and positron spectra at TeV energies and offer general comments on their interpretation. Subsequently, we focus on the study of the stochastic fluctuations and a secondary model for the positron excess.
}%

\section{Introduction}

In recent years, measurements of cosmic ray (CR) electrons and positrons have received heightened attention. There are four space experiments that are currently online and have provided measurements: AMS-02, CALET, DAMPE and \textit{Fermi}-LAT. The quality of data is exquisite, yet there are discrepancies that will need to be worked out. 

There are a number of reasons why CR electrons and positrons are interesting and provide information beyond the nuclear component which is dominating both in terms of number and energy density. First, the positron excess discovered some 15 years ago by the PAMELA satellite experiment~\cite{PAMELA:2008gwm} still has not been unambiguously be clarified. The most popular explanations so far have been self-annihilation of particle dark matter, pulsars/pulsar wind nebulae as well as production and subsequent acceleration of secondaries in old supernova remnants. While there exist complementary observations and constraints from other channels, better measurements of CR electrons and positrons likely hold the key to identifying the source of the positron excess.

Second, the study of the transport of CR electrons and positrons gives additional, complementary information for studying the transport in the Galaxy. Due to their small masses electrons and positrons suffer significant radiative losses while nuclei above a few GeV virtually do not lose energy at all. The spatial and spectral distribution of electrons and positrons therefore differ from those of nuclei and a solid understanding and successful modelling of both components is required when searching for the sources of CRs. 

Third and relatedly, the strong energy losses of TeV electrons and positrons limit the range of sources that can contribute to the observed fluxes at Earth to young and nearby objects. Consequently, the relevant number of sources should be fairly small, such that we can hope to identify spectral features from individual sources~\cite{Malyshev:2009tw,Mertsch:2018bqd}. Therefore, CR electrons and positrons could hold the key to identifying the sources of CRs.

\begin{figure}
\centering
\includegraphics[scale=0.7,trim={2cm 0 0 0.5cm}, clip=true]{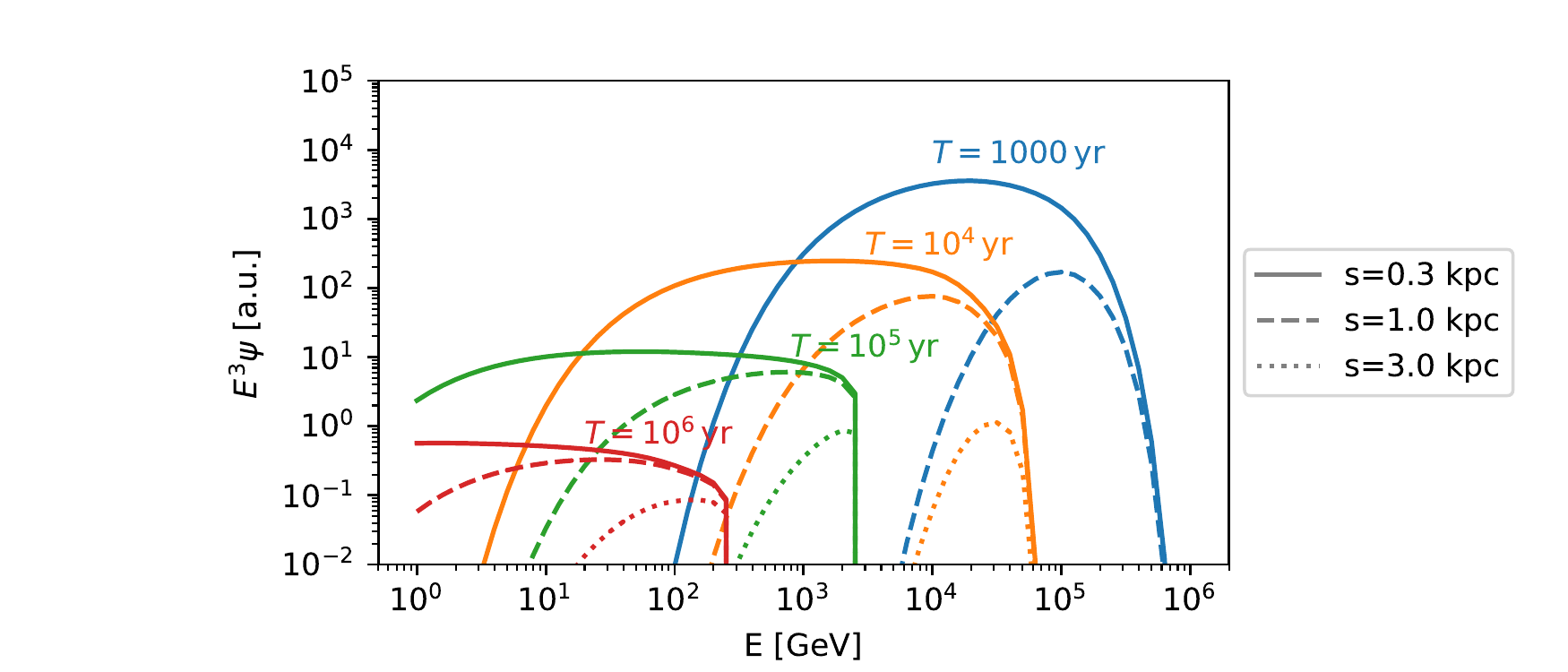}
\caption{Green's function of the transport eq.~\eqref{eqn:transport} with a source spectrum $Q(E_0) \propto E_0^{-2.2} \exp[-E_0/(10^5 \, \text{GeV})]$. The diffusion coefficient is of the form $\kappa(E) \propto E^{0.6}$, the half-height of the CR transport volume $z_{\text{max}} = 3 \, \text{kpc}$ and the energy losses are modelled as described in Ref.~\protect\cite{Mertsch:2018bqd}. The distances of the source are indicated by the line styles, the ages by the colours.}
\label{fig1}
\end{figure}

The last point requires some additional explanation. Under some simplifying assumptions, the transport of CR electrons and positrons in the Galaxy can be modelled by the following transport equation~\cite{Berezinsky:1990qxi,Gabici:2019jvz},
\begin{align}
\frac{\partial \psi}{\partial t} - \vec{\nabla} \cdot \kappa \cdot \vec{\nabla} \psi + \frac{\partial}{\partial E} \left( b(E) \psi \right) = q(\vec{r}, E, t) \, .
\label{eqn:transport}
\end{align}
Here, $\psi$ denotes the spectral density of electrons and positrons, $\kappa$ is the isotropic diffusion coefficient, $b(E) = \dd E / \dd t$ parametrises the rate of radiative losses and $q$ is the rate of injection of electron and positron density from sources. The Green's function for this partial differential equation is defined as the solution for a point-like, burst-like source at position $\vec{r}_0$ and time $t_0$ with a given spectrum $Q(E)$, that is 
\begin{align*}
q(\vec{r}, E, t) = \delta(\vec{r} - \vec{r}_0) \delta(t - t_0) Q(E) \, ,
\end{align*}
Examples of the Green's function for various distances and ages are shown in Fig.~\ref{fig1}. 
The flux from the whole population of Galactic sources is then just the sum of the Green's functions for all individual sources. At low energies, where the diffusion-loss-length is larger than the average source distance, individual point-like and burst-like sources can be approximated by a smooth source density on the right-hand side of eq.~\eqref{eqn:transport}. Consequently, the spectrum at these energies will be fairly smooth. At high energies, however, where only a few sources contribute, this would be a bad approximation. Instead, individual sources contribute their Green's functions and the spectrum will be rather bumpy. Identifying such features, individual sources can be identified.

\section{The all-electron flux}

\begin{figure}
\centering
\includegraphics[scale=0.8]{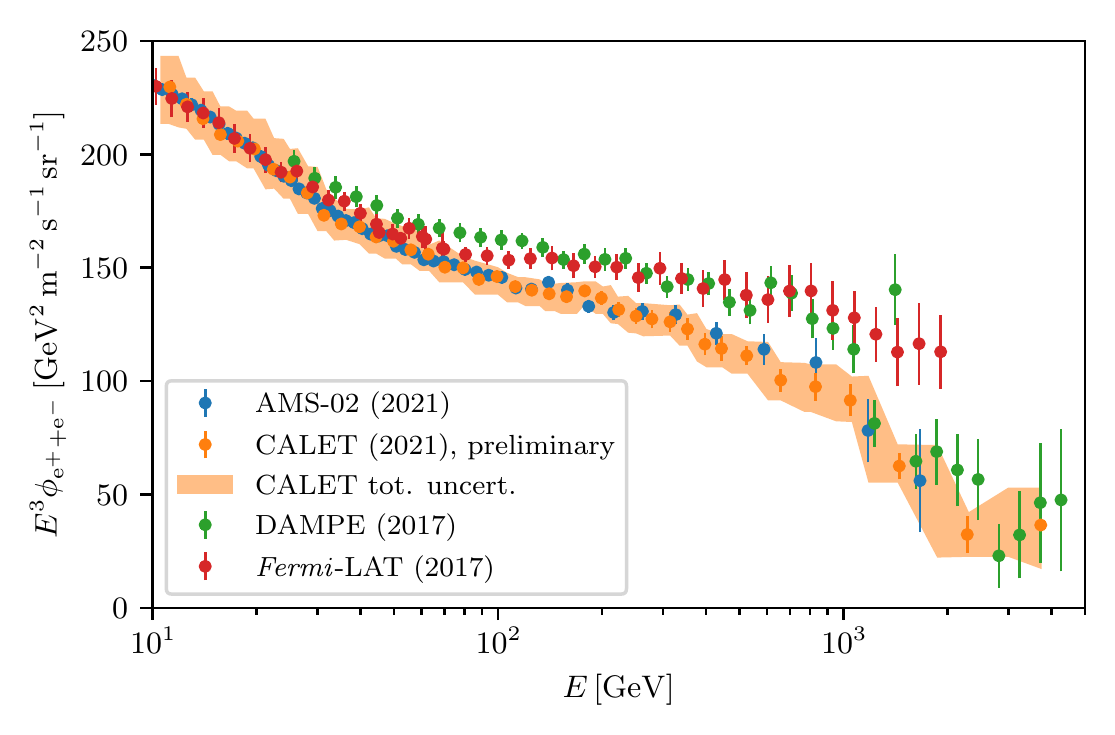}
\caption{All-electron flux measurements by AMS-02~\protect\cite{AMS:2021nhj}, CALET~\protect\cite{Torii:2021vjo}, DAMPE~\protect\cite{DAMPE:2017fbg} and \textit{Fermi}-LAT~\protect\cite{Fermi-LAT:2017bpc}.}
\label{fig2}
\end{figure}
The latest measurements of the all-electron flux from AMS-02~\cite{AMS:2021nhj}, CALET~\cite{Torii:2021vjo}, DAMPE~\cite{DAMPE:2017fbg} and \textit{Fermi}-LAT~\cite{Fermi-LAT:2017bpc} are compiled in Fig.~\ref{fig2}. Between $10 \, \text{GeV}$ and $\sim 30 \, \text{GeV}$ the data are well-described by a power law $E^{-3.25}$. At higher energies the spectrum flattens before it steepens significantly around $1 \, \text{TeV}$. However,  the various experiments do not agree on the detailed spectrum between $\sim 30 \, \text{GeV}$ and the cut-off: While DAMPE and \textit{Fermi}-LAT prefer a spectrum $\propto E^{-3.1}$, AMS-02 and CALET find a spectrum closer to $\propto E^{-3.15}$. While this might seem like a minor difference, given the small statistical and systematic uncertainties quoted, this disagreement is significant.

\begin{figure}
\centering
\includegraphics[scale=0.5]{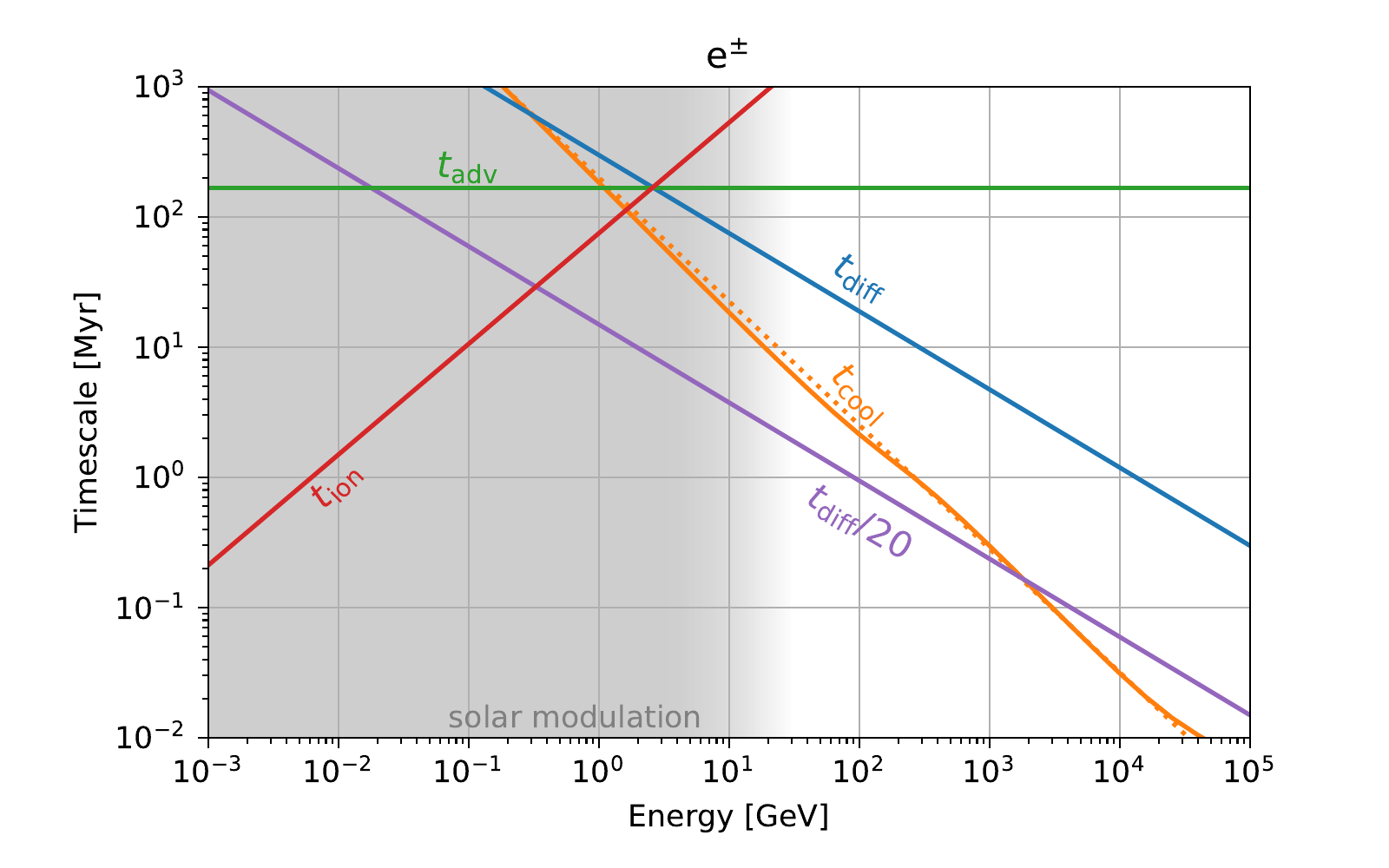}%
\caption{Time-scales relevant for the transport of CR electrons and positrons in the Galaxy. The transport parameters adopted are $\kappa(E) = 5 \times 10^{28} \, \text{cm}^2 \, \text{s}^{-1} (E/10 \, \text{GeV})^{1/3}$, $z_{\text{max}} = 5 \, \text{kpc}$; Klein-Nishina cross-section with energy density $\rho = \{ 0.26, 0.6, 0.6, 0.1 \} \, \text{eV} \, \text{cm}^{-3}$ for CMB, IR, opt, UV, characteristic temperatures of $\{20, 5 \times 10^3, 2 \times 10^4 \} \, \text{K}$ for IR, opt, UV as well as a $3 \, \mu\text{G}$ B-field; $n_{\text{H}} = 0.5 \, \text{cm}^{-3}$ (WIM) and $n_{\text{H}} = 0.5 \, \text{cm}^{-3}$ (WNM) and $100 \, \text{pc}$ wide gas disk. The advection speed assumed is $30 \, \text{km} \, \text{s}^{-1}$ and solar modulation is thought to become important below $\sim 10 \, \text{GeV}$.}
\label{fig3}
\end{figure}

As for the interpretation, it is useful to remind ourselves of the salient features expected for the all-electron spectrum if transport is described by the diffusion-loss eq.~\eqref{eqn:transport}: This can be best illustrated by the time-scales involved, see Fig.~\ref{fig3}. If there was a range in energy where diffusive losses were dominant, an $\propto E^{-\Gamma}$ source spectrum would be softened by diffusive losses with diffusivity $\kappa \propto E^{\delta}$ to an $E^{-\Gamma-\delta}$ ambient spectrum, in the same way as happens for CR nuclei. If cooling losses dominate, instead, it is an elementary exercise~\cite{Mertsch:2010fn} to show from eq.~\eqref{eqn:transport} that the ambient spectrum is $E^{-\Gamma-(\delta+1)/2}$. For instance, a source spectrum with $\Gamma = 2.3$ and a diffusion coefficient with $\delta = 0.6$ results in an ambient spectral index of $-\Gamma-(\delta+1)/2 = -3.1$, close to what is being measured below $\sim 1 \, \text{TeV}$. 

Of course, there is more interesting information in the spectral features, that is the hardening break around $30 \, \text{GeV}$ and the softening break at $\sim 1 \, \text{TeV}$. It has been suggested for instance, that the break at $30 \, \text{GeV}$ is the signature of the transition from the Thomson to the Klein-Nishina regime for the inverse Compton scattering that dominates the cooling of electrons and positrons at these energies~\cite{Evoli:2020ash}. Other authors disagree~\cite{DiMauro:2020cbn} and instead suggest a model with two different population of sources with different spectra contributing to the all-electron flux, e.g. supernova remnants at lower and pulsar wind nebulae at higher energies. This author has found that the strength and position of the spectral break from the Klein-Nishina suppression depends sensitively on the model for the interstellar radiation field. With the values adopted as in Fig.~\ref{fig3}, we find the break to take place at a few hundred GeV instead of a few tens of GeV.

The TeV break in the all-electron spectrum is probably even more contentious. A non-standard interpretation is that of a cooling break due to the transition between escape-dominated losses and cooling-dominated losses. Of course, this break--in fact a rather smooth transition--takes place at energies much lower than $1 \, \text{TeV}$ should the standard parameters be adopted. It has been suggested, however,~\cite{Lipari:2018usj} that the diffusive escape time has been overestimated by one to two orders of magnitude. In the right-hand panel of Fig.~\ref{fig3} we have indicated the resulting escape time and adjusted it in such a way that the transition would be taking place at about $1 \, \text{TeV}$. Interpreted in a diffusion model, such a short escape time would correspond to a CR diffusive volume of small half-height $z_{\text{max}}$. However, this seems to be in tension with preliminary data from the AMS-02 experiment on Beryllium isotope ratios.

\section{Source stochasticity}

As we have argued above, at high enough energies, the loss-length of CR electrons and positrons will be shorter than the mean distance between sources. Approximating the actual distribution of sources in space and time with a smoothly varying source density will then lead to significant errors. Instead, the fluxes at an arbitrary position will become very sensitive to the distances and ages of actual sources. We refer to this effect as source stochasticity. 
Where precisely this effect becomes important and what information can be extracted from measurement are important questions. 

In the literature, there is a spectrum of models that were suggested to deal with source stochasticity. Here, we identify three classes: 

\paragraph{Selective models.~\protect\cite{Kobayashi:2003kp,Recchia:2018jun,Fornieri:2020kch}} A limited set of usually nearby and young sources is modelled individually, while sources further afield are treated in the smooth approximation. It is easy to see that this can lead to spectral features beyond the maximum energy of the contribution from the smooth distribution. The question of how likely the considered contribution is to arise from a statistical model of sources is oftentimes not addressed. 

\paragraph{Catalogue models.~\protect\cite{Malyshev:2009tw,Manconi:2018azw,DiMauro:2020cbn}} These models usually start from a catalogue of specific sources, like pulsars and supernova remnants. While this removes some of the arbitrariness for the previous approach, it inherits some of the limitations of said catalogues, specifically selection biasses and incompleteness. Particularly worrisome is the absence of nearby but old sources which would not be active sources of radiation, but would still contribute to the flux of low-energy cosmic ray electrons. 

\paragraph{Probabilistic models.~\protect\cite{Blasi:2010de,Mertsch:2010fn,Mertsch:2018bqd,Cholis:2018izy,DiMauro:2020cbn,Evoli:2021ugn,Orusa:2021tts,Cholis:2021kqk}} Giving up constraints on the existence or non-existence of individual sources altogether, these models only make use of statistical information gained from surveys, for instance the density of sources as a function of galacto-centric radius. Moments of the flux distribution characterising the expectation value and the level of fluctuations can be computed. Alternatively, realisations of source distances and ages are drawn from this distribution in a Monte Carlo (MC) approach and the statistical moments are computed numerically. While this comes at the price of not being able to predict the actual realisation of the fluxes at Earth, the statistical analysis can be made very sound. In the following, we review the study of Mertsch (2018)~\cite{Mertsch:2018bqd}. 

Considering measurement of a CR flux in a finite number of energy bins, the fluxes at different energies can be combined into a flux vector. In a statistical ensemble, the flux in an energy bin is a random variable, and the vector of fluxes is a random vector. Previous studies have not considered the correlation between different energies. However, those correlations do exist and contain important information, which is encoded in the joint distribution of $f(\phi_1, \phi_2, \mathellipsis \phi_N)$. 

Unlike for the marginals, where the flux distribution can be estimated analytically, a direct computation of the joint distribution seems forbidding. Using Monte Carlo simulations, one might consider kernel density estimators, however, they will fail due to the curse of dimensionality for more than a handful of energy bins. Instead, we have recently employed a copula construction~\cite{Nagler:2016}, a method for constructing an approximate joint distribution that has gained currency in financial mathematics. This builds on Sklar's theorem which factorises an arbitrary $N$-dimensional joint distribution into a product of $N$ 1D probability distribution functions (PDFs) $f_i(\phi_i)$, the marginals, and the so-called copula, a function of the $N$ cumulative distribution functions (CDFs), $F_i(\phi_i)$. It can be shown that the copula can be decomposed into a nested product which contains only (conditional) bi-variate PDFs. Details of this construction can be found in Ref.~\cite{Mertsch:2018bqd}. 

\begin{figure}
\centering
\includegraphics[width=\textwidth,trim={0 0 0 0}, clip=true]{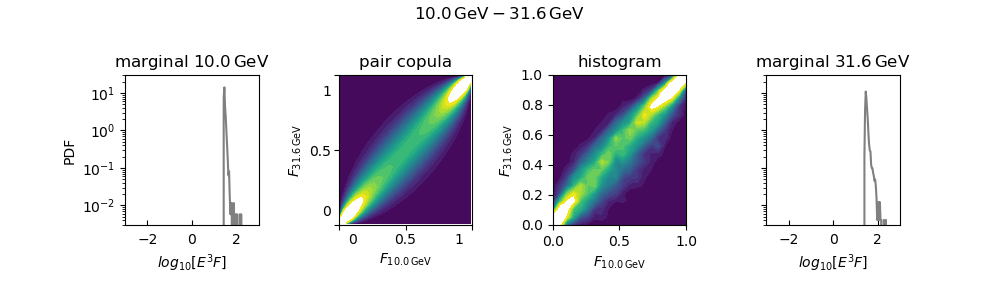}	
\includegraphics[width=\textwidth,trim={0 0 0 0}, clip=true]{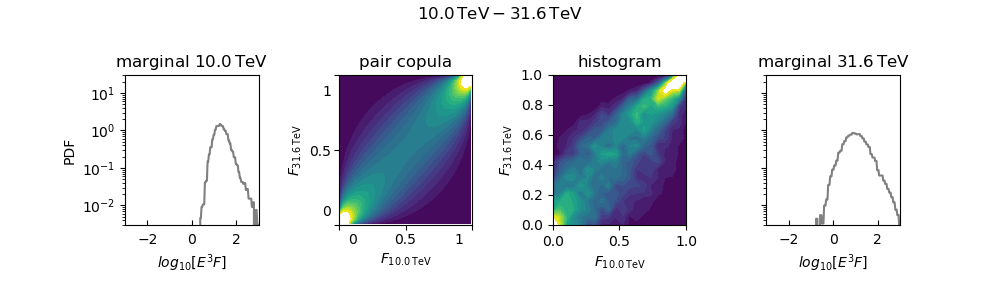}	
\caption{Variance and correlations in the flux distribution due to source stochasticity. \textbf{Top panel.} The left-most and right-most panel show the flux PDF estimated by the histograms of (logarithmic) flux at $10$ and $31.6 \, \text{GeV}$, respectively. The central panels compare the pair copular with the histogram of the empirical flux CDF at both energies. \textbf{Bottom panel.} The same as the top panel, but for $10$ and $31.6 \, \text{TeV}$.}
\label{fig4}
\end{figure}

We have parametrised the joint distribution of high-energy electron-positron fluxes with a pair copula construction and determined the free copula parameters by fitting to a large number of MC samples. A comparison between the pair-copulas which are functions of the CDFs and the histogram of empirical CDFs is shown in Fig.~\ref{fig4}. We have substituted the measurements by H.E.S.S. of the all-electron flux between a few hundred GeV and $\sim 20 \, \text{TeV}$ into this likelihood. The log-likelihood is the sum of contributions from the marginals, quantifying the fluctuations around the expectation value, and from the copula, quantifying the correlations between different energy bins. 

Both contributions resulting from the H.E.S.S. data can be compared with the distribution of log-likelihood contributions obtained with the simulated fluxes. This comparison depends of course on the model parameters assumed, for instance on the source rate. With a canonical rate of supernova remnants, that is $10^{-2} \, \text{yr}^{-1}$, the contribution from the marginals has a probability to exceed (PTE) of $0.4$, but the copula contribution has $\text{PTE} = 0.998$. Put differently, the observed flux has less correlations than would be expected from the simulations. If we decrease the source rate by one order of magnitude while increasing the power of each individual source by one order of magnitude, the expectation value of the flux is largely unchanged if the cut-off energy is adjusted, yet the level of fluctuations is much increased. We note that a reduced source rate can be considered an approximation for a class of sources where CR accelerators do not occur independently in space and time, but where several sources appear in conjunction, an example being a number of supernova remnants appearing in super bubbles. It appears that this allows for a much better agreement between observations and simulations, that is $\text{PTE} = 0.466$ (marginals) and $\text{PTE} = 0.298$ (copula). We conclude that our analysis thus supports a scenario where sources take place with a certain spatial and temporal coherence. 

\begin{figure}
\centering
\includegraphics[scale=0.5,trim={0 0.2cm 0 1.2cm},clip=True]{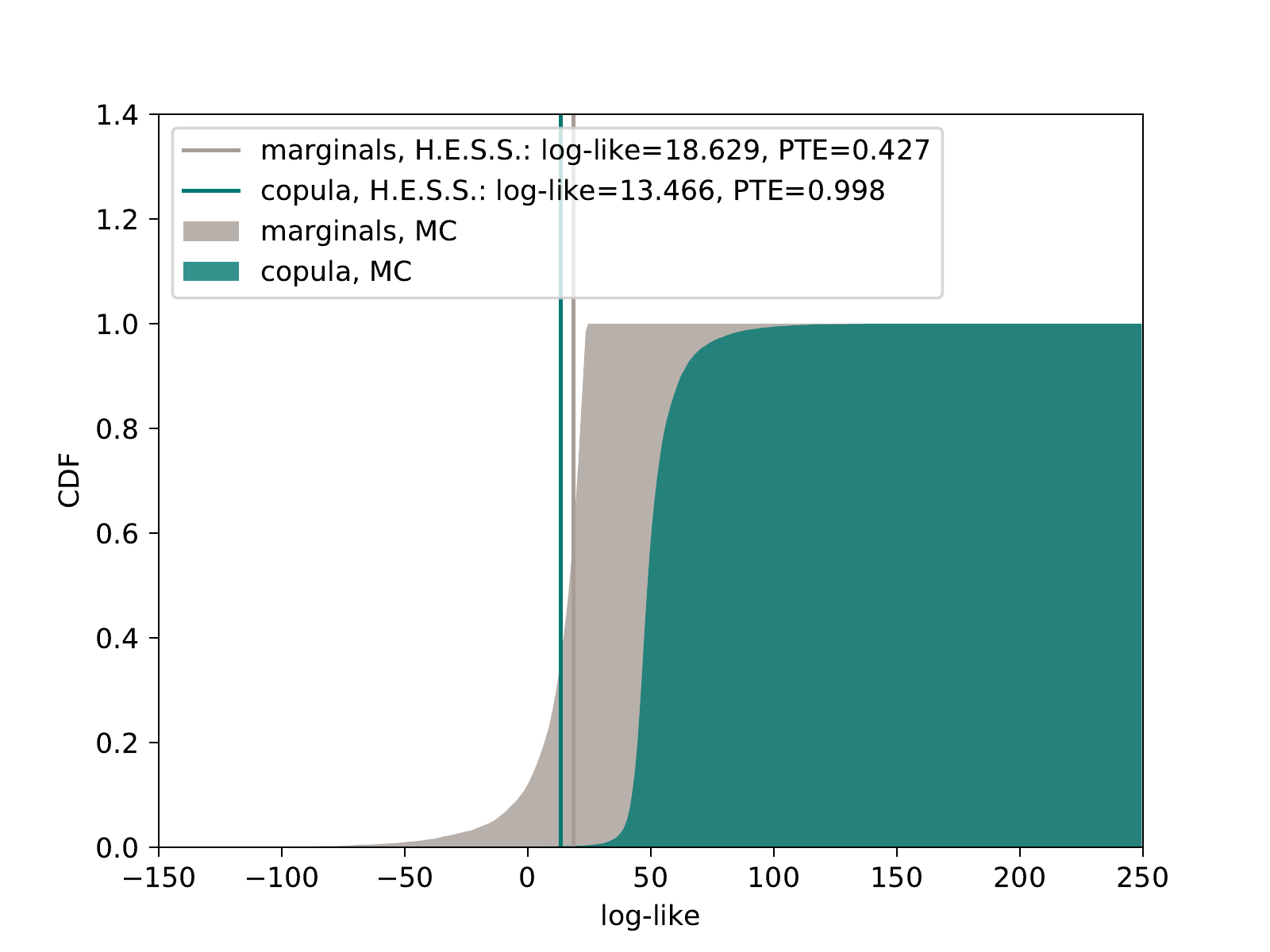}%
\includegraphics[scale=0.5,trim={0 0.2cm 0 1.2cm},clip=True]{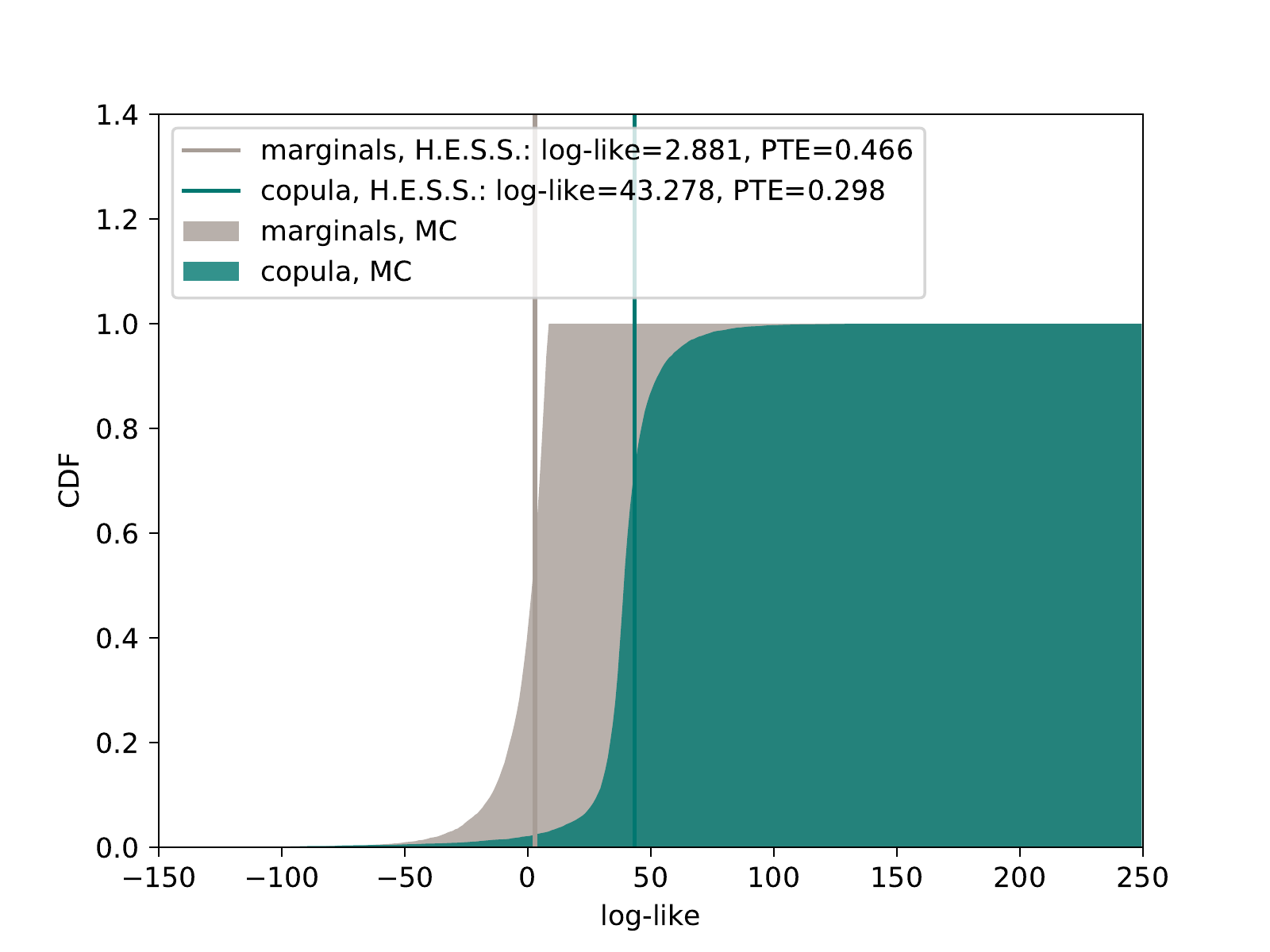}%
\caption{CDFs for the marginal and copula contribution to the log-likelihood, both for the MC simulations and for the H.E.S.S. data. \textbf{Left panel:} For the canonical source rate of $10^{-2} \, \text{yr}^{-1}$, the copula contribution for the data is much lower than what is expected from simulations. \textbf{Right panel:} For a decreased source rate of $10^{-3} \, \text{yr}^{-1}$, the copula contribution for the data agrees well with the simulations.}
\label{fig5}
\end{figure}

\section{Positrons}

\begin{figure}[!b]
\centering
\includegraphics[width=0.6\textwidth]{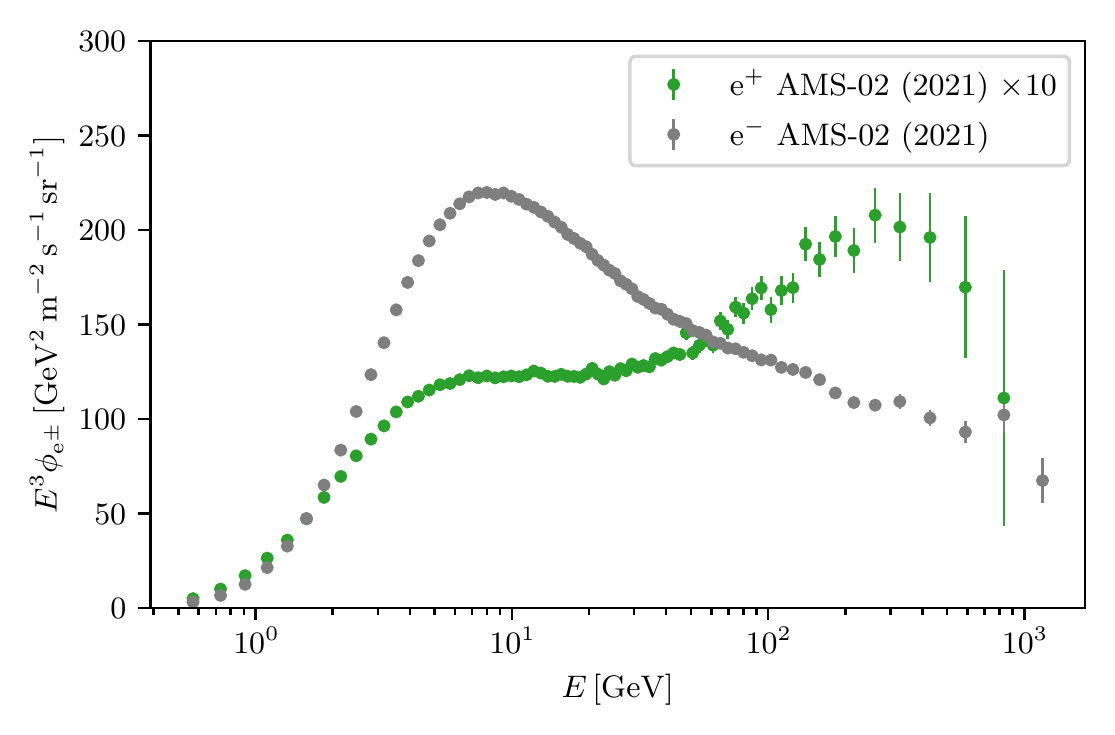}
\caption{Electron and positron spectra as measured by AMS-02~\protect\cite{AMS:2021nhj}.}
\label{fig6}
\end{figure}

\begin{figure}[!thb]
\centering
\includegraphics[scale=1]{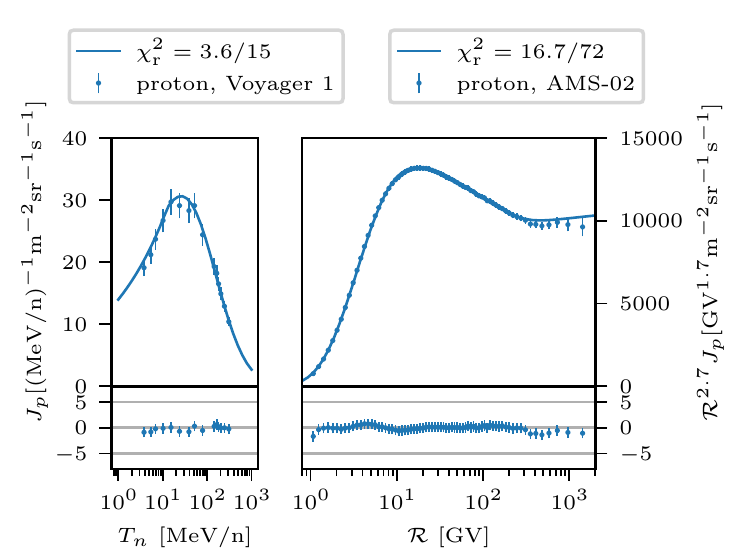}
\includegraphics[scale=1]{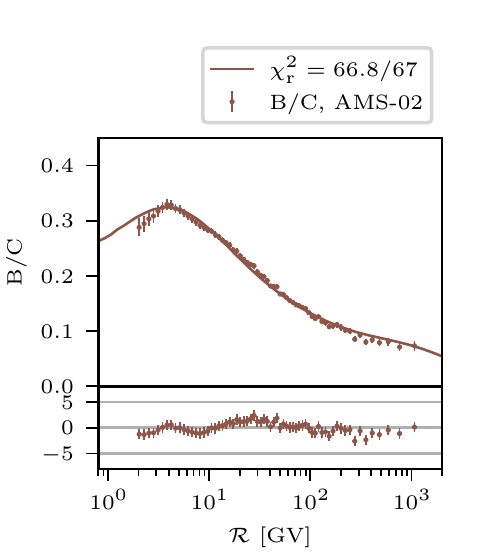}
\includegraphics[scale=1]{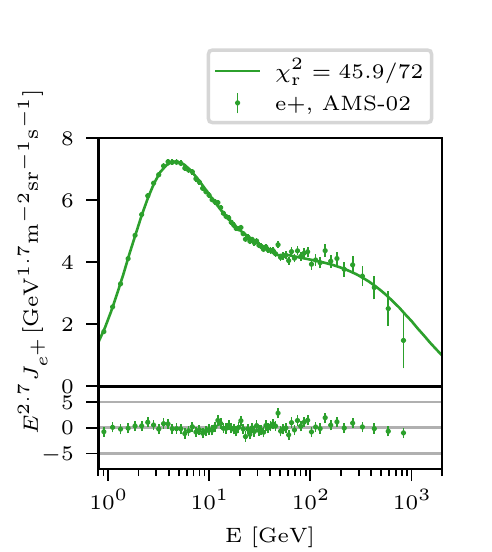}
\includegraphics[scale=1]{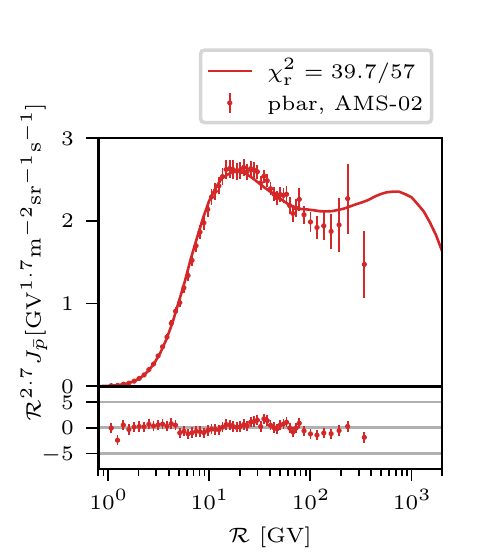}
\caption{Comparison between proton flux (top left), boron-to-carbon ratio (top right), positron flux (bottom left) and antiproton flux (bottom right) as predicted from the model with acceleration of secondaries model~\protect\cite{Mertsch:2020ldv} and as measured by AMS-02~\protect\cite{AMS:2015tnn,AMS:2018tbl,AMS:2019rhg,AMS:2016oqu}.}
\label{fig7}
\end{figure}

So far, we have only discussed the interpretation of the all-electron flux. 
Spectrometric experiments like PAMELA and AMS-02, however, can distinguish electrons from positrons and have thus provided separate spectra. We show the measurement from AMS-02 in Fig.~\ref{fig6}; note that we have upscaled the positron flux by a factor 10 in order to be able to compare the spectral shapes. Both fluxes have been multiplied with energy cubed and so it is apparent that the electron flux is slightly softer than $E^{-3}$ above $\sim 10 \, \text{GeV}$ while the positron flux is harder. 

A positron spectrum harder than $E^{-3}$ is incompatible with secondary production in the interstellar medium (ISM), at least in the conventional scenarios of Galactic transport. Therefore, a new source of CR positrons is required. The most exciting possibility would have been non-baryonic dark matter producing electrons and positrons through self-annihilation. With the energies involved, masses at the weak scale seemed preferred, however, the necessary cross-sections needed to be significantly enhanced compared to the cross-section implied by thermal freeze-out scenarios. Unfortunately, this possibility is now severely constrained by observations of gamma-rays, CR antiprotons and the CMB. Pulsars or rather pulsar wind nebulae (PWNe) are a popular, astrophysical explanation for the hard positron spectrum. PWNe must be contributing at some level and other contributions at this meeting have covered such models, e.g.\ Ref.~\cite{Orusa:2021tts}. Here, we present another astrophysical expalantion, that is old supernova remnants. 

In fact, the lore that the positron excess cannot be due to secondary positrons is based on a misunderstanding. It is usually said that the positron's primaries, e.g.\ protons, spend relatively short amounts of time inside the sources, for instance supernova remnants, compared to the transport time in the Galaxy. Even if the gas density inside the source is enhanced compared to the ISM density, the grammage accumulated inside the sources is thus estimated to be negligible. This argument ignores, however, that charged secondaries like positrons can undergo shock acceleration much like their primaries, but will attain a spectrum generically harder than the primaries. For instance, assuming a Bohm-like scaling of the diffusion coefficient for the source, $\kappa(p) \propto p$, the secondary spectra will be harder than the primary ones by one power of momentum. This will lead to a harder spectrum of positrons after propagation than in the standard scenario where production and acceleration of secondaries in sources is ignored. Of course, other charged secondaries like anti-protons or boron are equally affected, which provides for a nice test of this scenario. 

We have recently revisited this issue and fitted our model to the data from AMS-02. Particular care was given to testing various parametrisations of the cross-sections of positron and anti-proton production. We have managed to find a very good fit to CR protons, helium, carbon, oxygen, boron, nitrogen, positrons and antiprotons. We compare our model fluxes with data for a selection of species in Fig.~\ref{fig7}. Further details can be found in Ref.~\cite{Mertsch:2020ldv}.


\end{document}